\documentclass[aps,pra,twocolumn,showpacs]{revtex4}
\usepackage{graphicx,amssymb,amsmath}
\def\be{\begin{equation}}
\def\ee{\end{equation}}
\def\ba{\begin{eqnarray}}
\def\ea{\end{eqnarray}}

\begin{document}
\title{Tunneling Time in the Landau-Zener Model}
\author{Yue Yan and Biao Wu}
\affiliation{Institute of Physics, Chinese Academy of Sciences,
Beijing 100190, China}
\date{November 10, 2008}

\begin{abstract}
 We give a general definition for the tunneling time in
 the Landau-Zener model.  This definition allows us to
 compute numerically the Landau-Zener tunneling time at any sweeping rate
 without ambiguity. We have also obtained analytical results
 in both the adiabatic limit and the sudden limit. Whenever applicable,
 our results are compared to previous results and they
 are in good agreement.
\pacs{03.65.Ge, 32.80.Bx, 33.80.Be, 34.70.+e}
\end{abstract}

\maketitle
\section{INTRODUCTION}
Tunneling is one of many fundamental quantum processes that have no
classical counterparts. It exists ubiquitously in quantum systems
and is a key to understanding many quantum
phenomena\cite{LZ7,LZ302,LZ301,LZ30,LL1,LL2,LL3,LZ33,LZ32,LZy,LZy1,app}.
Discussions on tunneling can be found in all textbooks on quantum
mechanics. However, these discussions are mainly focused on the
probability of tunneling from one quantum state to another or from
one side of a potential barrier to the other. In contrast, there are only a few
extensive and in-depth discussions in literature
on another aspect of tunneling, the time of tunneling, that is, how
long a tunneling process lasts\cite{PZ1,PZ2,PZ3,PZ4,PZ5,LZ3}. This
disparity is partly caused by the difficulty to define properly
tunneling times in many situations.

The difficulty of having a proper definition for tunneling time has
its root in the wave or probabilistic nature of quantum mechanics,
and is best demonstrated in the  example of a wave packet tunneling
through a potential barrier. In this case, one would intuitively
define the tunneling time as the time spent by the peak (or
centroid) of the wave packet under the barrier.  However, as pointed
out by Landauer and Martin\cite{PZ51}, if this definition was used,
a packet could leave the barrier before entering it. To overcome
this difficulty, many different definitions have been suggested, and
no clear consensus has been reached so far\cite{PZ0}.

The focus of this study is on the tunneling time in the Landau-Zener
(LZ) model\cite{LZ1,LZ2}. This system is much simpler than the wave
packet and barrier system. Nevertheless, a proper definition of
tunneling time in this model is still missing in spite of the
studies in the past by  many authors. Mullen \emph{et al.}
\cite{LZ8} discussed the LZ tunneling time in the diabatic basis for
the two limiting cases, the adabatic limit and the sudden limit.
They found that for large $\Delta$ (the adiabatic limit), the
tunneling time scales with  $\Delta/\alpha$ and for small $\Delta$
(the sudden limit), the tunneling time  is about
$\sqrt{\hbar/\alpha}$. $\Delta$ is the minimal energy gap between
the two eigenstates in the LZ model and $\alpha$ is the sweeping
rate. However, Mullen \emph{et al.} did not give a general
definition for the tunneling time. Vitanov\cite{LZ5} has given a
much more thorough study on the LZ tunneling time. He obtained
analytical results for the LZ tunneling time in both diabatic and
adiabatic bases. Furthermore, Vitanov tried to give a general
definition for the tunneling time. However, his definition fails in
certain cases, in particular, the adiabatic limit.

In this paper, we give a general definition for the tunneling time
in the LZ model.  We show both analytically and numerically that
this definition yields reasonable results at any sweeping rate in
both adiabatic basis and diabatic basis. With this general
definition, we are able to reproduce the previous results obtained
by Mullen \emph{et al.}\cite{LZ8} and Vitanov\cite{LZ5}.
Furthermore, we are able to compute analytically the tunneling time
in the adiabatic basis at the adiabatic limit, which is given by
$2\sqrt{2^{1/3}-1}\Delta/\alpha$. This, to our best knowledge, has
not been obtained before.

Besides it theoretical significance,  our work has also potential
applications. In the Monte-Carlo simulation of quantum tunneling in
molecular magnets, the authors in Ref.\cite{LZ4} have used an
empirical formula for the LZ tunneling time. This formula, given by
$\sqrt{\Delta^2/\alpha^2+2\hbar/\alpha}$ and interpolating the two
limiting results in Ref.\cite{LZ8}, is not well founded. With this
newly-proposed definition, one no longer needs this empirical
formula to do the Monte-Carlo simulation.

 We note here that it is important to study the tunneling time in both
 the diabatic basis and the adiabatic basis. When the LZ model is applied
 to the case of the tunneling of spin under a sweeping magnetic field\cite{LZ4},
 the diabatic basis is a better choice. When the LZ model is applied to
 the tunneling between Bloch bands under a constant force, it is better
 to use the adiabatic basis\cite{LZ31}.

Our paper is organized as follows. In Sec.{\ref{sec0}}, we introduce
our definition of the tunneling time in the LZ model and we analyze
the effectiveness of our definition. In Sec. {\ref{sec1}}, we
present our results of the tunneling times, which include the
analytical results at the adiabatic limit and the sudden limit and
the numerical results for the general case. Our results are given
both in the diabatic basis and the adiabatic basis. In the last
section, we discuss our results and conclude.

\section{DEFINITION OF the Tunneling TIME}
\label{sec0}
The LZ model is a two-level system and is described by \cite{LZ1,LZ2}
\begin{equation}
 i\hbar\frac{d}{dt}\bigg(\begin{matrix}a(t) \\b(t) \end{matrix}\bigg)
 =H(\gamma)\bigg(\begin{matrix}a(t) \\b(t) \end{matrix}\bigg)\,,
\label{y2}
\end{equation}
where
\begin{equation}
H(\gamma)=\bigg(\begin{matrix}\gamma/2 & \Delta/2\\ \Delta/2 &
-\gamma/2\end{matrix}\bigg) \label{our1}\,,
\end{equation}
with $\gamma=\alpha t$ changing with time linearly. $\alpha$ is
usually called sweeping rate.  The tunneling behavior in the LZ
model can be described in two different bases, diabatic basis and
adiabatic basis\cite{LZ5}. In the diabatic basis, we use
$P_d(t)=|b(t) |^2$ to describe the LZ tunneling dynamics. In the
adiabatic basis, we use
\begin{equation}
P_a(t)=|a^*(t)a(\gamma(t))+b^*(t)b(\gamma(t))|^2\,,
\end{equation}
where $(a(\gamma),b(\gamma))$ is the instantaneous eigenstate of
$H(\gamma)$.  When our discussion is independent of the basis, we
will remove the subscript and simply use $P$ for both $P_d$ and
$P_a$.

\begin{figure}
\includegraphics[width=8cm]{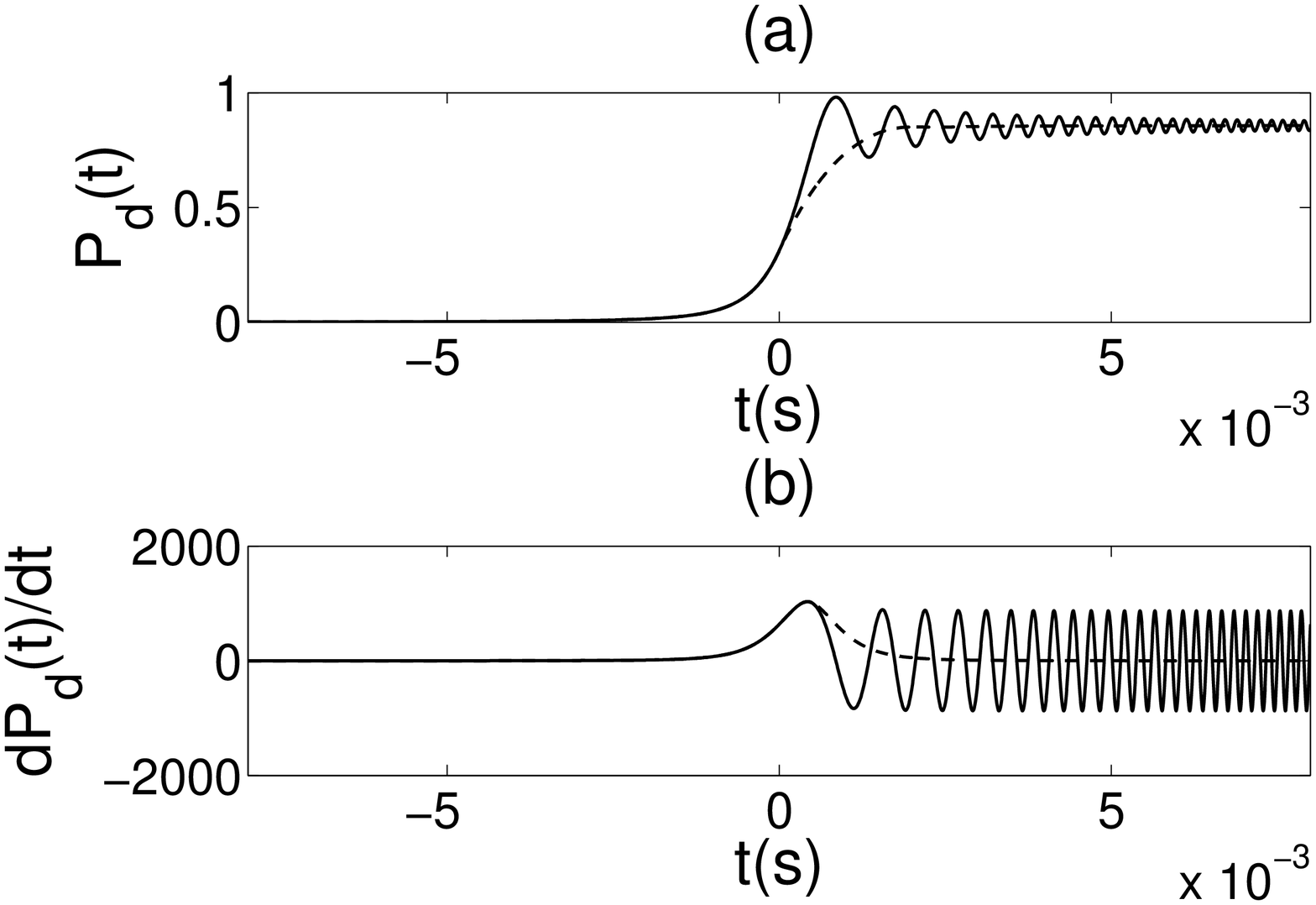}
\caption{(a)The solid line is the time evolution of
the probability function $P_d(t)$ and the dashed line is its step-like function fit.
(b) The time derivative of the two functions in (a) and $\eta=0.2565$.}
 \label{definition}
\end{figure}

\begin{figure}
\includegraphics[width=8cm]{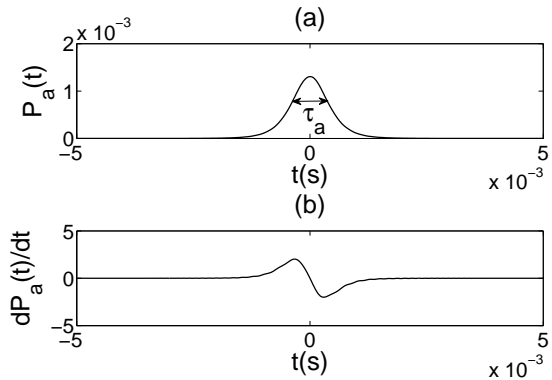}
\caption{Time evolution of tunneling probability $P_a(t)$ and its
time derivative $dP_a(t)/dt$ at the adiabatic limit.  (a) The time
evolution of the tunneling probability $P_a(t)$; (b) the function of
$dP_a(t)/dt$, $\eta=0.10$.}\label{ap4}
\end{figure}

The time evolution of the probability function $P(t)$ can be found
by numerically solving Eq.(\ref{y2}).  A typical  result of $P(t)$
is shown in Fig.\ref{definition}, where we see a sharp transition
occurs around $t=0$ and is followed by decaying oscillations. This
observation suggests an intuitive (or natural) definition for the LZ
tunneling time. One may first fit the $P$ curve with a smooth
step-like function (dashed line in Fig.\ref{definition}(a), and then
define the half-width of its time derivative (dashed line in
Fig.\ref{definition}(b)) as the tunneling time. However, like its
counterpart in the wave packet and barrier system, this intuitive
definition of tunneling time fails because of its two shortcomings.
First, there are numerous methods to find the fitting step-like
function in Fig.\ref{definition}(a); there is no obvious criterion
by which one method is better than the other. Secondly, at the
adiabatic limit, the $P$ curve looks drastically different from the
typical case in Fig.\ref{definition}(a). As shown in Fig.\ref{ap4},
the $P$ is a single-peaked function; one really has to be
far-stretched to fit it with a step-like function. These two
drawbacks show that this intuitive definition of the tunneling time
based curve-fitting is not a good choice. One has to find an
alternative.

\begin{figure}
\includegraphics[width=8cm]{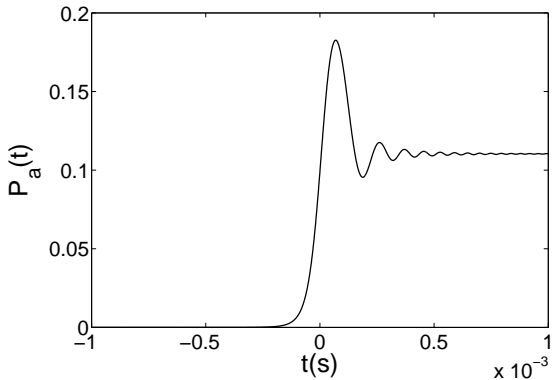}
\caption{Time evolution of the tunneling probability in the adiabatic basis. $\eta=1.425$.}
\label{yy0}
\end{figure}

In Ref.\cite{LZ5}, Vitanov introduced a general definition for the
LZ tunneling time. His definition is
\begin{equation}
\tau=\frac{P(\infty)}{P^{'}(0)} \label{NV1}\,,
\end{equation}
where  the time derivative value $P^{'}(0)$ at $t=0$ is used to
represent  the  rate of the transition around $t=0$. For the typical
time evolution of the probability function $P(t)$ shown in
Fig.\ref{definition},  this definition works well.  However,
Vitanov's definition fails at the adiabatic limit like the intuitive
definition as we shall see later. In fact, Vitanov's definition does
not work in the adiabatic basis in general. A typical $P$ curve in
the adiabatic basis is shown in Fig.\ref{yy0}, where $P^\prime (0)$
clearly over-represents the transition rate.

We overcome these difficulties and find a general definition of the
tunneling time for the LZ model. In the definition, we first find a
$t'<0$ such that
\begin{equation}
P(t')=\frac{1}{2}P_{max}\,, \label{t0}
\end{equation}
where $P_{max}$ is the maximum value of $P$ when $t\leq 0$. Usually,
$P_{max}=P(t=0)$.  The above condition will be called  half-width condition
from now  on for ease of reference.  We then  introduce two more variables
\begin{align}
\label{our2}
S_1=&\int_{-\infty} ^0 \frac{d}{dt}P(t)dt=P(0)\,,\\
S_2=&\int_0 ^\infty \frac{d}{dt}P(t)dt=P(\infty)-P(0)\,,
\end{align}
which are the left ($t<0$) area and right ($t>0$) area of the $dP/dt$
curve, respectively. With these defined variables,  we define the
tunneling time as
\begin{equation}
\tau=|t'|(1+\Big|\frac{S_2}{S_1}\Big|)\,. \label{our}
\end{equation}

Three quick remarks. {\it i}) The three variables in the definition
can be computed without any ambiguity. {\it ii}) For the typical
case shown in Fig.\ref{definition}, we have $S_1\approx S_2$ and,
therefore, $\tau=2|t'|$, which is in agreement of the ``intuitive"
definition that we discussed before. {\it iii}) The absolute value
is used because $S_2$ can be negative in certain cases, for example,
the case in Fig.\ref{ap4}. The reason of the appearance of negative
$S_2$ is that the transition around $t=0$ over-shoots the overall
transition $P(\infty)-P(-\infty)$ and the system needs to spend some
time to ``wind back".

In the following, we shall apply our definition and compute the tunneling
times in the LZ model.  Both analytical and numerical approaches will be used.
The analytical approach is used for two limiting cases, the adiabatic limit and
the sudden limit. For the LZ model, we can introduce a ``quickness"
parameter
\begin{equation}
\eta\equiv\frac{2\hbar\alpha}{\Delta^2}\,.
\end{equation}
The adiabatic limit is $\eta\ll 1$ while $\eta\gg 1$ corresponds
to the sudden limit.
For a general case,  we have to resort to the numerical method. We first solve
numerically the equation of motion Eq.(\ref{y2}), then compute the tunneling probability
function $P(t)$, and finally find the tunneling time with our definition in Eq.(\ref{our}).
In our computation, we use $\Delta=1.2\times10^{-7}k_b$, where $k_b$ is
Boltzmann constant, which is a typical value in molecular magnet\cite{app}.

The tunneling time will be computed in both
the adiabatic basis and the diabatic basis. For clarity, we shall
use $\tau_d$ for the tunneling time in the diabatic basis and $\tau_a$
for the tunneling time in the adiabatic basis.

\section{Tunneling Times In The Diabatic Basis}
\label{sec1}
We first consider the diabatic basis and follow it with the discussion on
the adiabatic basis in the next section.
\subsection{Analytical results}
At the adiabatic limit ($\eta\ll1$), according to Vitanov\cite{LZ5}
\begin{equation}
P_d(t)\approx\frac{1}{2}+\frac{\alpha
t}{2\sqrt{\alpha^2t^2+\Delta^2}}\,, \label{wu1}
\end{equation}
the variable $t'$ can be obtained from  the half-width condition
\begin{equation}
P_d(t)|_{t=t'}=\frac{1}{2}(P_d(t\leq0))_{max}=\frac{1}{4}\,.
\end{equation}
The result is
\begin{equation}
t'=-\frac{\sqrt{3}}{3}\frac{\Delta}{\alpha}\label{y10}\,.
\end{equation}
Since we have $S_1=S_2$ for the tunneling curve Eq.(\ref{wu1}),
the tunneling time with our definition of Eq.(\ref{our}) is
\begin{equation}
\tau^a_d=\frac{2\sqrt{3}}{3}\frac{\Delta}{\alpha}\label{y20}\,,
\end{equation}
which agrees well with the result of Mullen \emph{et al.}\cite{LZ8}.

At the sudden limit $\eta\gg1$,  it is beneficial to take a
transformation
\begin{align}
a(t)&=\tilde{a}(t)\exp(-i\frac{\alpha t^2}{4\hbar})\,,\\
b(t)&=\tilde{b}(t)\exp(i\frac{\alpha t^2}{4\hbar})\,,
\end{align}
for the LZ model.  As a result, the diagonal terms in the
Hamiltonian are transformed away and we can expand $\tilde{a}(t)$
and $\tilde{b}(t)$ in powers of $\eta$ (effectively, $\Delta$) \cite{LZ8}. For the initial
condition $a(-\infty)=1$ and $b(-\infty)=0$, we obtain
\begin{align}
\label{y91}
a(t)&=[1+\sum_{k=1}^{\infty}(-1)^k\frac{1}{(2\eta)^k}a_{2k}(y)]\exp(-i\frac{y^2}{4})\,,\quad\\
\label{y9}
b(t)&=[\sum_{k=1}^{\infty}(-1)^{k+1}\frac{1}{i(2\eta)^{k/2}}b_{2k-1}(y)]\exp(i\frac{y^2}{4})\,,\quad
\end{align}
where
\begin{widetext}
\begin{align}
a_n(y)&=\int_{-\infty}^y \exp(ix_1^2/2)\int_{-\infty}^{x_1}
\exp(-ix_2^2/2)\dots\int_{-\infty}^{x_{n-1}}
\exp[(-1)^{n+1}ix_n^2/2]dx_1dx_2\dots dx_{n-1}dx_n\,,\quad\\
b_n(y)&=\int_{-\infty}^y \exp(-ix_1^2/2)\int_{-\infty}^{x_1}
\exp(ix_2^2/2)\dots \int_{-\infty}^{x_{n-1}}
\exp[(-1)^{n}ix_n^2/2]dx_1dx_2\dots
dx_{n-1}dx_n\,,\quad\label{y0}\raggedleft
\end{align}
\end{widetext}
with $y= t/(\hbar/\alpha)^{1/2}$.  At the sudden limit,   it is
sufficient to keep  Eq.(\ref{y9}) to the lowest order of $1/\eta$.
Consequently, we obtain
\begin{align}
\label{sudden1}
P_d(t)&=|b(t)|^2\approx\bigg|\frac{1}{\sqrt{2\eta}}\int_{-\infty}^y\exp(-i\frac{x^2}{2})dx\bigg|^2\\\notag
&=\frac{1}{2\eta}\bigg|-\frac{\sqrt{2\pi}}{2}\exp(i\frac{3\pi}{4})+\int_0^y\exp(-i\frac{x^2}{2})dx\bigg|^2\\\notag
&=\frac{\pi}{2\eta}\Big\{\big[\frac{1}{2}+C(\frac{y}{\sqrt{\pi}})\big]^2+
\big[\frac{1}{2}+S(\frac{y}{\sqrt{\pi}})\big]^2\Big\}^2\,,
\end{align}
where $C(y/\sqrt{\pi})$ and $S(y/\sqrt{\pi})$ are the Fresnel
integrals\cite{Fresnel}. One can prove that the maximum value of
$P_d$ for $t\le 0$ is at $t=0$, where $P_d=\pi/(4\eta)$. Thus with
the half-width condition
\begin{equation}
P_d(t')=\frac{1}{2}P_d(0)=\frac{\pi}{8\eta}\label{sudden2}\,,
\end{equation}
we find numerically that  $t'\approx-0.6241\sqrt{\frac{\hbar}{\alpha}}$.
According to Refs.\cite{LZ1,LZ2}, we have
\begin{equation}
P_d(\infty)=1-\exp(-\frac{\pi}{\eta})\,.
\end{equation}
Therefore, at the sudden limit ( $\eta\gg1$), we have
\begin{equation}
\Big|\frac{S_2}{S_1}\Big|=\frac{P_d(\infty)-P_d(0)}{P_d(0)}\approx 3\,.
\end{equation}
Based on our definition in Eq. (\ref{our}),  the tunneling time is
\begin{equation}
\tau^s_d\approx4|t'|=2.4964\sqrt{\frac{\hbar}{\alpha}},\label{sudden3}
\end{equation}
which agrees well with the result of Mullen \emph{et al.}\cite{LZ8}.

\begin{figure}
\includegraphics[width=8cm]{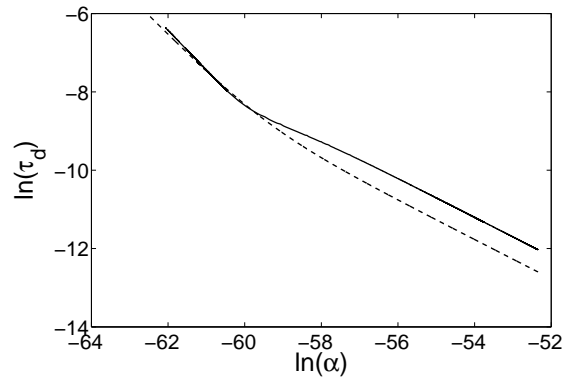}
\caption{Tunneling times $\tau_d$ in the diabatic basis. The solid
line is our numerical results and
the dashed line is the empirical formula
$\tau_d=\sqrt{\Delta^2/\alpha^2+2\hbar/\alpha}$
used in Ref.\cite{LZ4}. $\alpha$ is in the unit of $\frac{\Delta^2}{2\hbar}$.}
\label{yy3}
\end{figure}

\subsection{Numerical results}

\begin{figure}
\includegraphics[width=8cm]{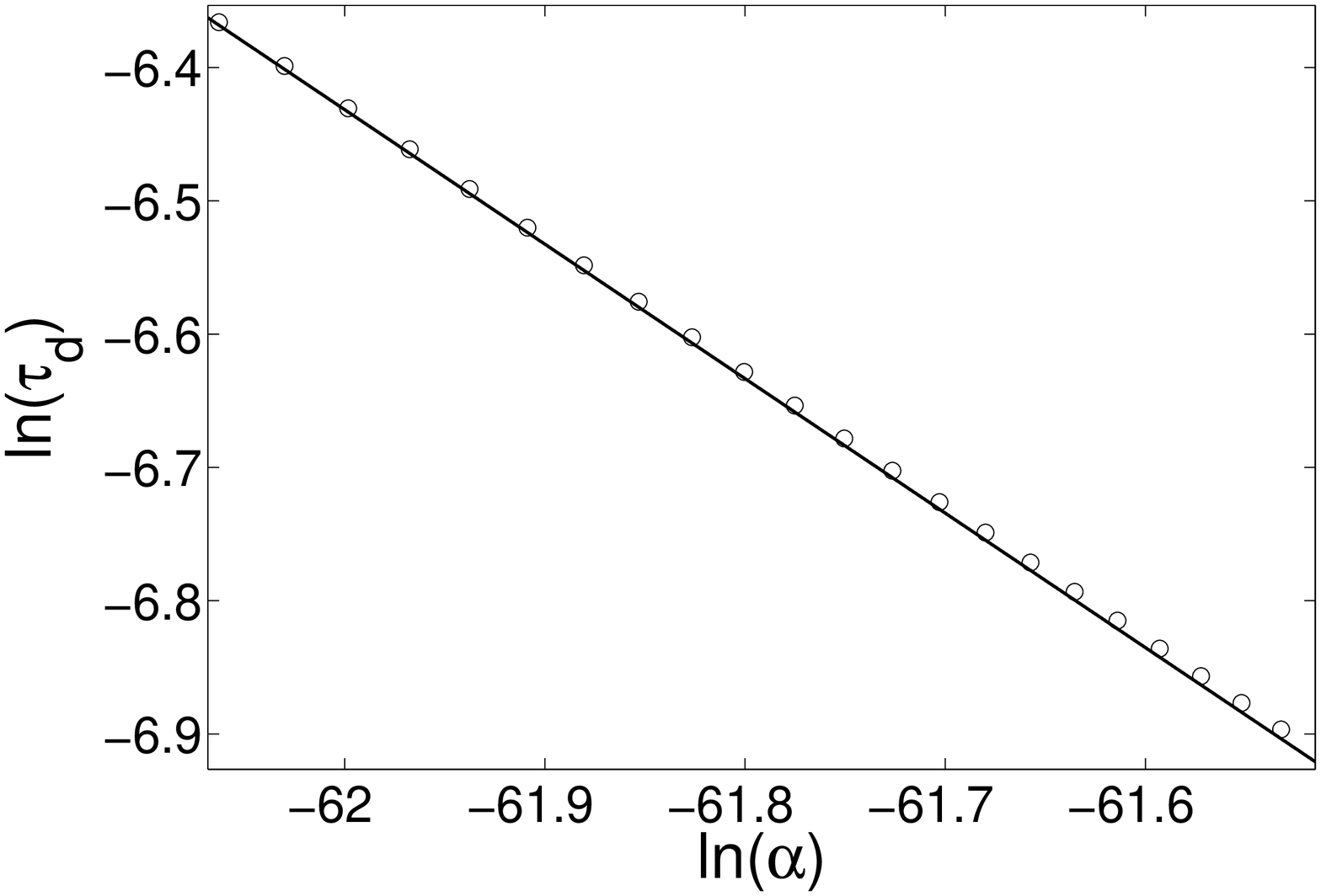}
\caption{Tunneling time $\tau_d$ at the adiabatic limit in the
diabatic basis. The circles are the theoretical results given in
Eq.(\ref{y20}) and the solid line is the numerical results. $\alpha$
is in the unit of $\frac{\Delta^2}{2\hbar}$.}\label{yy4}
\end{figure}
\begin{figure}
\includegraphics[width=8cm]{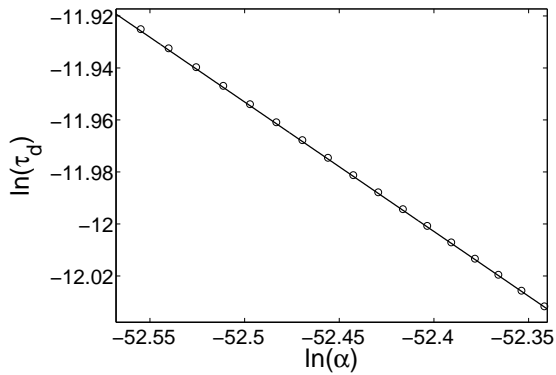}
\caption{Tunneling time  $\tau_d$ at the sudden limit in the
diabatic basis. The circles are the theoretical results given in
Eq.(\ref{sudden3}) and the solid line is the numerical results.
$\alpha$ is in the unit of $\frac{\Delta^2}{2\hbar}$.}\label{yy5}
\end{figure}
Our numerical results of the tunneling time $\tau_d$ in the diabatic
basis is plotted in the log-log scale in Fig.\ref{yy3}.  In the
figure, we see that the results for the two limiting cases are
connected by a  smooth kink. We have also compared these results
with the empirical relation
$\tau_d=\sqrt{\Delta^2/\alpha^2+2\hbar/\alpha}$ that was used in
Ref.\cite{LZ4}; the agreement is quite good.

We have amplified the results at the adiabatic and the sudden limits
and plotted them in Fig.\ref{yy4}. and Fig.\ref{yy5}, respectively.
In these two figures, we have also compared them to the analytical
results and the agreement is excellent.

\section{Tunneling Times  In The Adiabatic  Basis}
As mentioned already, when one applies the LZ model to describe the tunneling
between Bloch bands,   it is more  convenient to use the adiabatic basis. It
turns out that the results in the adiabatic basis are quite different from the ones
in the diabatic basis.

\subsection{Analytical results}
As in the case of the diabatic basis, at the adiabatic
limit($\eta\ll1$), the tunneling probability function $P_a$ in the
adiabatic basis has been found by Vitanov \cite{LZ5},
\begin{equation}
P_a(t)\approx\frac{\alpha^2\hbar^2\Delta^2}{4(\alpha^2t^2+\Delta^2)^3}\,.
\end{equation}
With the half-width condition,
\begin{equation}
P_a(t)|_{t=t'}=\frac{1}{2}(P_a(t\leq0))_{max}=\frac{\alpha^2\hbar^2}{8\Delta^4}\,,
\end{equation}
we find that
\begin{equation}
t'=-\sqrt{2^{1/3}-1}\frac{\Delta}{\alpha}\,.
\label{y15}
\end{equation}
Thus, based on our definition in Eq.(\ref{our}), the tunneling time is
\begin{equation}
\tau^a_a=2\sqrt{2^{1/3}-1}\frac{\Delta}{\alpha}\label{our3}\,,
\end{equation}
which is very similar to the tunneling time Eq.(\ref{y20}) in the
diabatic basis.  In contrast,  according to Vitanov's definition
Eq.(\ref{NV1}),  the tunneling time is\cite{LZ5}
\begin{equation}
\zeta_a^a=\frac{\sqrt{2}\Delta}{\sqrt{\alpha\hbar}}\exp(-\frac{\pi\Delta^2}{4\alpha\hbar})\,.
\end{equation}
At the adiabatic limit ($\alpha\rightarrow0$), the tunneling time $\zeta_a^a$  tends to
be zero. This result contradicts with the physical reality, the tunneling time should be
very long at the adiabatic limit.  Thus, Vitanov's definition does not work  in this case.
Alternatively, our result  in Eq. (\ref{our3}) can be viewed as the first successful attempt
to find the tunneling time at the adiabatic limit in the adiabatic basis.

We  next consider the sudden limit  ($\eta\gg1$). In terms of $y=t/\sqrt{\hbar/\alpha}$,
the instantaneous eigenstates of the Hamiltonian (\ref{our1}) are
\begin{eqnarray}
\Phi(y)=\begin{pmatrix}c_1(y)\\c_2(y)\end{pmatrix}=
\begin{pmatrix}[\frac{1}{2}(1+\frac{y}{\sqrt{(y^2+2/\eta)}})]^{1/2}\\
[\frac{1}{2}(1-\frac{y}{\sqrt{(y^2+2/\eta)}})]^{1/2}\end{pmatrix}\,.\label{ex}
\end{eqnarray}
According to Eq. (\ref{y91}) and Eq. (\ref{y9}), we can obtain the tunneling probability up to
the first order of $1/\eta$
\begin{eqnarray}
P_a(y)\approx\bigg|{\begin{pmatrix}c_1(y)&c_2(y)\end{pmatrix}\begin{pmatrix}
\exp(-i\frac{y^2}{4})\\b_{1}(y)\exp(i\frac{y^2}{4})\end{pmatrix}}\bigg|^2\label{yy41}\,.
\end{eqnarray}
With Eq.(\ref{y0}), we arrive at
\begin{widetext}
\begin{equation}
P_a(y)\approx\frac{1}{2}+\frac{1}{2}\frac{y}{\sqrt{y^2+\frac{2}{\eta}}}-\frac{1}{\sqrt{\eta(\eta y^2+2)}}
\cos\frac{y^2}{2}\int_{-\infty}^y\sin{\frac{x^2}{2}}dx+\frac{1}{2\eta}(\frac{1}{2}-\frac{1}{2}
\frac{y}{\sqrt{y^2+\frac{2}{\eta}}})\,.
\label{wu2}
\end{equation}
\end{widetext}
It is quite obvious that the last two terms of the above equation is much
smaller than the first two terms. Finally, we obtain
\begin{equation}
P_a(t)\approx\frac{1}{2}+\frac{1}{2}\frac{\alpha t}{\sqrt{\alpha^2 t^2+\Delta^2}}\,,\\\label{wu3}
\end{equation}
which is surprisingly identical to the result at the adiabatic limit
in the diabatic basis (see Eq.(\ref{wu1})). As a result, we can
similarly obtain  the tunneling time
\begin{equation}
\tau^s_a=\frac{2\sqrt{3}}{3}\frac{\Delta}{\alpha}\,,
\label{sudden4}
\end{equation}
which agrees very well with Vitanov's result\cite{LZ5}.

\subsection{Numerical results}

\begin{figure}[!h]
\includegraphics[width=8cm]{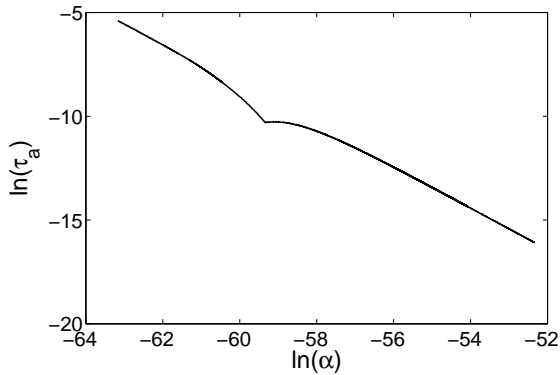}
\caption{Tunneling time $\tau_a$ in the adiabatic basis, $\alpha$ is in the unit of $\frac{\Delta^2}{2\hbar}$.}
\label{yy6}
\end{figure}
\begin{figure}[!h]
\includegraphics[width=8cm]{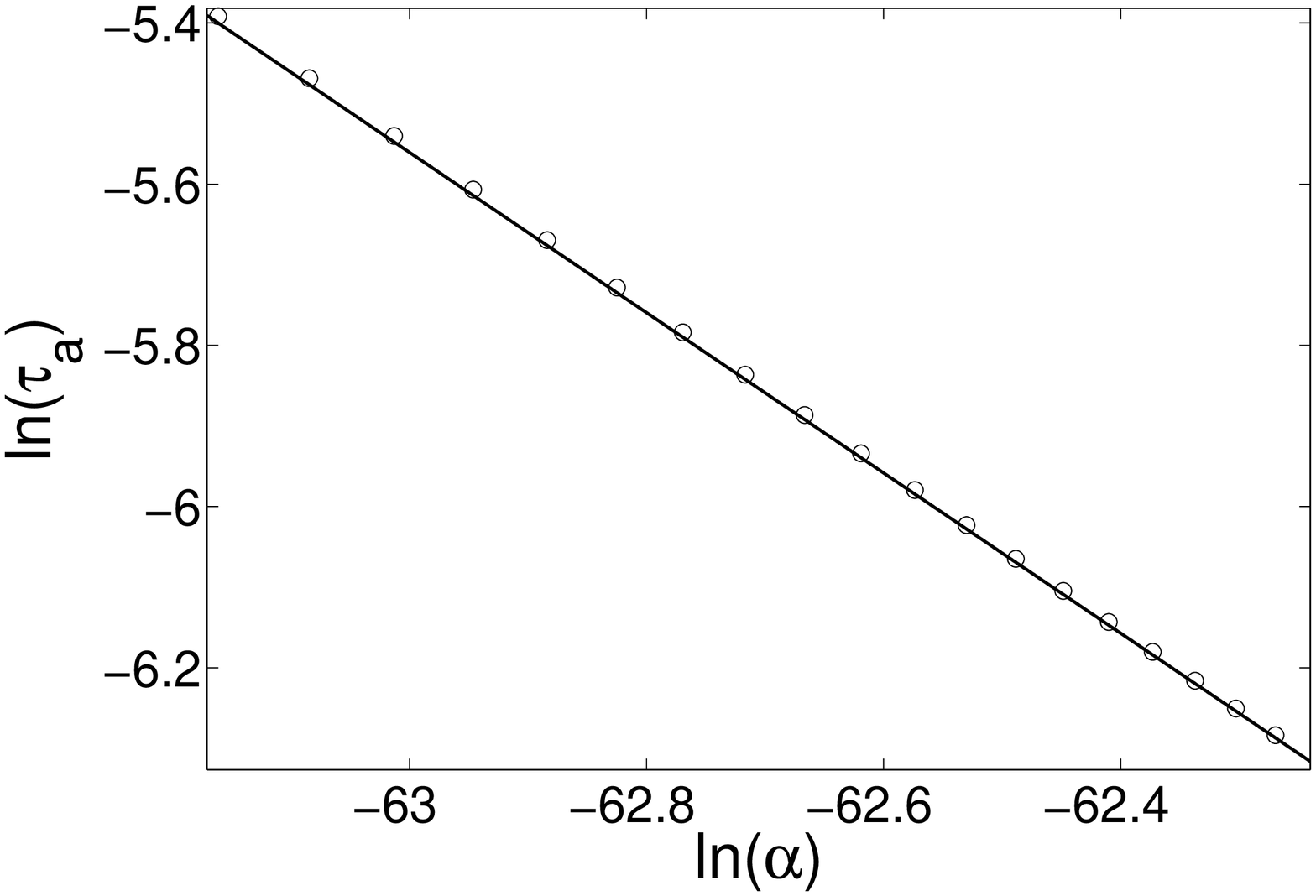}
\caption{Tunneling time $\tau_a$ at the adiabatic limit in the
adiabatic basis. The circles are the theoretical results given by
Eq.(\ref{our3}); the solid line is the numerical results. $\alpha$
is in the unit of $\frac{\Delta^2}{2\hbar}$.}\label{yy7}
\end{figure}
\begin{figure}[!h]
\includegraphics[width=8cm]{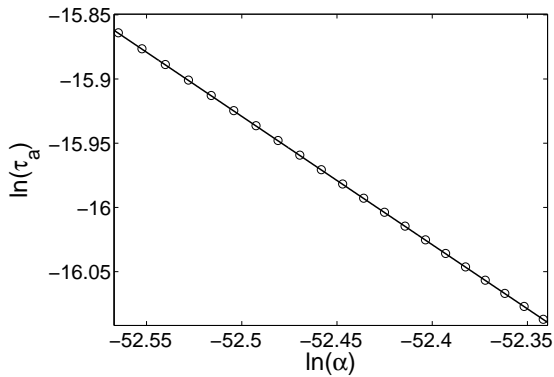}
\caption{Tunneling time $\tau_a$ at the sudden limit in the
adiabatic basis. The circles are the theoretical  results given by
Eq.(\ref{sudden4}); the solid line is the numerical results.
$\alpha$ is in the unit of $\frac{\Delta^2}{2\hbar}$.}\label{yy8}
\end{figure}

In the adiabatic basis, the numerical results of the  tunneling time
are shown in Fig.\ref{yy6}.  We see that the results in the two
limiting cases are also connected by a kink. However, this kink is
not as smooth as the kink in the diabatic basis; the first
derivative of the tunneling time with respect to $\alpha$ is not
continuous. The numerical results for the two limits, the adiabatic
limit and the sudden limit, are plotted and compared to the
theoretical results in Fig.\ref{yy7} and  Fig.\ref{yy8},
respectively. Again, we find an excellent agreement.

\section{DISCUSSION AND CONCLUSION}
In the above, we have obtained analytical results for the tunneling
times in the LZ model at two different limits and in two different
bases. They are, respectively, $t_d^a$, $t_d^s$, $t_a^a$, and
$t_a^s$.  We have found that  only $t_d^s$ is proportional to
$\sqrt{\hbar/\alpha}$ while the rest of the three tunneling times
all scale as $\Delta/\alpha$. It is not hard to understand why
$t_d^s$, the tunneling time at the sudden limit and in the diabatic
basis, does not scale as $\Delta/\alpha$. The effect of $\Delta$ is
to couple the two bare states, $(1,0)$ and $(0,1)$. which serve as
the base vectors in the diabatic basis. At the sudden limit, the
system changes very fast and its wave function remains almost
unchanged. As a result, the system does not feel the effect of
$\Delta$. It is also not hard to understand that the two tunneling
times at the adiabatic limit,  $t_a^a$ and $t_d^a$,  scale as
$\Delta/\alpha$. At the adiabatic limit, the effect of $\Delta$ is
fully felt by system and gets reflected in the tunneling time.

The most puzzling is $t_a^s$, the tunneling time at the sudden limit
in the adiabatic basis. Unlike the other tunneling time $t_d^s$ at
the sudden limit,  it is proportional to $\Delta/\alpha$. Moreover,
its corresponding probability function $P_a(t)$ described by
Eq.(\ref{wu3}) is surprisingly identical to the probability function
$P_d(t)$ in Eq.(\ref{wu1}), which is at the adiabatic limit in the
diabatic basis.  To understand this, we have to look into the
details of the evolution. At the adiabatic limit, the system follows
its instantaneous eigenstate as demanded by the quantum adiabatic
theorem\cite{qat}. $P_d(t)$ in Eq.(\ref{wu1}) is obtained by
projecting this instantaneous eigenstate to the bare state $(0,1)$.
At the sudden limit, the wave function of the system changes little
and remains in the bare state $(1,0)$. However, in the adiabatic
basis, this wave function needs to be projected to the instantaneous
eigenstate to obtain $P_a(t)$ described by Eq.(\ref{wu3}).  As we
know, projecting a bare state to an instantaneous eigenstate is the
identical to projecting the same instantaneous eigenstate to the
same bare state. This explains why the probability function $P_a(t)$
in  Eq.(\ref{wu3})  is the same as $P_d(t)$ in Eq.(\ref{wu1}).
Consequently, this also explains why $t_a^s$ scales as
$\Delta/\alpha$.

In sum, we have presented a general definition of the tunneling time
for the Landau-Zener model. We have shown that this definition works
for any sweeping rate and can be used for the numerical computation
of  the tunneling time without any ambiguity. In particular, we have
obtained analytical results for the two limiting cases, the
adiabatic limit and the sudden limit. We have not only reproduced
known results but also found the tunneling time at the adiabatic
limit in the adiabatic basis, which has not been found before to our
knowledge.

\section{Acknowledgements}
This work was supported by the NSF of China (10504040, 10825417) and the 973 project of
China (2005CB724500, 2006CB921400).

\end{document}